\documentstyle[12pt]{article}

\newcommand{\be}{\begin{equation}} 
\newcommand{\ee}{\end{equation}} 
\newcommand{\bea}{\begin{eqnarray}} 
\newcommand{\eea}{\end{eqnarray}} 
 
\newcommand{\p}[1]{(\ref{#1})} 
  
\begin{document}
\begin{flushright}
hep-th/9704133\\
April 1997 
\end{flushright}
\centerline{\large\bf N=4 supersymmetric integrable systems}
\vskip0.6cm
\centerline{{\large E.A. Ivanov}}
\vskip.3cm
\centerline{\it Bogoliubov Laboratory of Theoretical Physics, JINR,}
\centerline{\it 141 980, Dubna, Russian Federation}
\vskip.3cm
\centerline{\it Talk at the International Seminar} 
\centerline{\it SUPERSYMMETRY AND QUANTUM FIELD THEORY}
\centerline{\it dedicated to the memory of Dmitrij Volkov} 
\centerline{\it Kharkov (Ukraine) January 5-7, 1997} 
\vskip.5cm

\begin{abstract}{\small
\noindent I give an overview of recent progress in constructing the KdV, 
mKdV and NLS type hierarchies with extended $N=4$ supersymmetry.}
\end{abstract}
\vskip.5cm

\noindent{\bf 1. Introduction.} It is widely believed nowadays
that the ultimate theory of all fundamental forces is by no means 
a standard field theory, but rather that of extended objects like 
superstrings and supermembranes. Supersymmetry, the concept 
pioneered by D.V. Volkov \cite{volkov}, will surely be 
one of the key-stones of this future theory. Affine and $W$ 
algebras and superalgebras are also expected to be necessary 
ingredients of the underlying symmetry structure of this 
theory as they naturally come out as the world-sheet (or world-volume) 
gauge symmetries of extended objects. 

The WZW, Liouville - Toda and KdV type $2D$ integrable systems
are intimately related to these symmetries. They are encountered and 
proved to be of high relevancy in a plenty of problems of modern 
mathematical physics (see, e.g., \cite{rev}): 
in non-perturbative $2D$ gravity and related matrix models, in the 
geometric approaches to strings, superstrings and supermembranes, 
in Seiberg-Witten 
non-perturbative approach to supersymmetric Yang-Mills theory, etc. KdV, 
mKdV, NLS and KP type hierarchies of evolution equations exhibit a 
remarkable relationship with conformal, affine and $W$ algebras: the 
latter provide a hamiltonian structure for the former \cite{2ham}. 
Supersymmetric extensions of these hierarchies and their interplay 
with superaffine and superconformal ($W$) algebras were under intense 
study for the last decade [4-10].  

The subject of my talk is the KdV type integrable hierarchies 
with extended $N=4$ supersymmetry. While a lot was known about $N=1$ and 
$N=2$ extensions, until recently it remained unclear whether consistent 
higher $N$ hierarchies of this kind exist. Only $N=4$ extensions of 
some exactly solvable Lorentz-covariant $2D$ systems, Liouville and WZW 
models, were known \cite{{IK},{IKL}}. Seeking higher $N$ hierarchies 
is of extreme interest, in particular, because such systems can be relevant  
to the program of ``grand-unification'' of all known 
hierarchies: apparently unrelated lower supersymmetry (and purely bosonic) 
integrable systems can turn out to be various reductions of a single 
higher $N$ supersymmetric system. This phenomenon can be already seen 
while passing from $N=1$ KdV hierarchy \cite{{Mat1},{Mat2}} to the 
$N=2$ ones \cite{{Mat3},{Mat4}}. The latter naturally 
incorporate both KdV and mKdV hierarchies in the bosonic sector. 

In a series of papers [13-15] the first example of 
KdV system with higher supersymmetry, $N=4$ KdV hierarchy, was constructed 
and analyzed, both in a manifestly supersymmetric $N=4$ superfield form 
\cite{DI}
and via $N=2$ superfields \cite{DIK}. It was found to be 
bi-hamiltonian, to have ``small'' $N=4$ SCA \cite{Ad} as the second 
hamiltonian 
structure and to possess two different Lax formulations in terms 
of $N=2$ super pseudo-differential operators \cite{{DG},{IK1}}. 
A remarkable interplay between integrability of this system and 
breaking of the global 
automorphism $SU(2)$ symmetry of $N=4$ supersymmetry  was revealed: 
it is integrable 
only provided this  $SU(2)$ is explicitly broken and the square of 
$SU(2)$ breaking parameter is proportional to the inverse of the central 
charge of $N=4$ SCA. It encompasses two different $N=2$ KdV hierarchies, 
the $a=4$ and $a=-2$ ones, as its two non-equivalent consistent reductions.
Later it was found that ``small'' $N=4$ SCA provides a hamiltonian 
structure for one more integrable hierarchy which is an extension of the 
$a=-2$, $N=2$ KdV hierarchy and possesses only $N=2$ global 
supersymmetry \cite{DGI}. Recently \cite{IKT}, a simplest $N=4$ 
supersymmetric affine hierarchy was constructed. It is defined on $N=2$ 
extension of the bosonic affine algebra $\widehat{sl(2)\oplus u(1)}$ and 
underlies $N=4$ KdV hierarchy much like ordinary mKdV hierarchy underlies 
(via a Miura map) the KdV hierarchy. It seems that any $N=2$ affine algebra 
or superalgebra admitting a quaternionic structure exhibits hidden $N=4$ 
supersymmetry and hence can give rise to $N=4$ supersymmetric hierarchies. 
This provides a general clue to constructing and classifying such 
hierarchies.

\vspace{0.2cm}
\noindent{\bf 2. KdV example.} To explain the basic idea of how 
to construct KdV type hierarchies via relating them to some 
infinite-dimensional algebras as second hamiltonian structure, let us 
start with the text-book KdV example. 

As was shown in \cite{2ham}, the KdV equation
\begin{equation}\label{a1}
\dot u =-u{}'''+ 6uu{}'
\end{equation}
can be treated as a hamiltonian system,
$$
\dot u = \left\{ u,{\cal H}_3\right\} \quad ,
$$
with the hamiltonian and the Poisson brackets defined by
\begin{equation}\label{a2}
{\cal H}_3 = \frac{1}{2} \int dx \; u^2(x) \quad , \;\;\;
\{ u(x),u(y) \} = \left[ -\partial^3+
  4u\partial+ 2 u{}' \right] \delta (x-y) \quad .
\end{equation}
For the Fourier modes of $u(x)$,
\begin{equation}\label{a5}
u(x)  =  {6\over c}\sum_n \mbox{exp}(-inx)L_n -{1\over 4} \quad,
\end{equation}
the Poisson brackets in (\ref{a2}) imply the structure relations of
the Virasoro algebra
\begin{equation}
i\left\{ L_n,L_m \right\} =  (n-m)L_{n+m}+{c\over 12}(n^3-n)
\delta_{n+m,0}\;.
   \label{a4}
\end{equation}
So, the  definition (\ref{a2}) means that the
density of the KdV hamiltonian ${\cal H}_3$ is the square of
a conformal stress-tensor $u(x)$ obeying the Virasoro algebra
(\ref{a2}), (\ref{a4}). One says that the Virasoro algebra provides 
the second hamiltonian (or Poisson) structure for the KdV equation 
(historically, first hamiltonian formulation of KdV hierarchy was based 
upon a linear Poisson algebra, and the latter is referred to as the 
first hamiltonian structure). The higher order conserved quantities 
of the KdV equation can be regarded as the hamiltonians which generate, 
through the Poisson brackets (\ref{a2}), next equations from 
the KdV hierarchy.

\vspace{0.2cm}
\noindent{\bf 3. N=1,2 KdVs.}
The same idea was applied for constructing $N=1$ and
$N=2$ superextensions of the KdV equation [5-9].
They were related in an
analogous way, via the second hamiltonian structure, to
$N=1$ and $N=2$ SCAs. In the $N=1$ case the basic object 
is $N=1$ stress-tensor, the spin $3/2$ fermionic $N=1$ superfield 
\be
\Phi(t,x,\theta) = \xi(t,x) + \theta u(t,x).  \label{n1}
\ee
It comprises the spin $3/2$ fermionic current $\xi$ and spin $2$ 
stress-tensor $u$ which generate, via appropriate PBs, the classical 
$N=1$ SCA. The most general $N=1$ supersymmetric dimension $3$ hamiltonian 
reads 
\be
{\cal H}_3 = \int dx d\theta\; \Phi D\Phi\;,\;\; D ={\partial \over 
\partial \theta} 
+\theta {\partial \over \partial x}\;, \; D^2 = \partial_x\;, 
\ee
and $N=1$ KdV equation is the equation defining evolution of $\Phi$ with 
respect to ${\cal H}_3$
\be 
\dot \Phi = \left\{ \Phi, {\cal H}_3\right\}\;,
\ee
with the superfield PB structure amounting to $N=1$ SCA.

In the $N=2$ case one deals with the $N=2$ stress-tensor which 
is a spin $1$ $N=2$ superfield 
\be 
J(t,x,\theta, \bar \theta) = j(t,x) + \theta \xi (t,x) + 
\bar \theta \bar \xi(t,x) + \theta \bar\theta u(t,x)\;, 
\ee
with the components being currents of $N=2$ SCA. Once again, the 
$N=2$ KdV equation can be defined as an evolution equation 
\be 
\dot J = \left\{ J, {\cal H}_3\right\} 
\ee
with respect to the most general $N=2$ supersymmetric 
dimension $3$ hamiltonian 
\be
{\cal H}_3 = \int dx d\theta d\bar \theta \left( J[D, \bar D]J + 
{a\over 6} J^3 \right)  \label{n2ham}
\ee
and the PB structure 
\be
\{ J(1), J(2) \} = \left( J\partial+ \partial J  + DJ \bar D + 
\bar D J D + \partial [D, \bar D] \right) \delta(1,2)\;, 
\label{N2pb} 
\ee
which generates $N=2$ SCA (we always choose the central charge 
equal to some number since on the classical level it can be fixed 
at will by proper rescalings of superfields and PBs). 
In these formulas
\be
\{D, D \} = 0\;, \;\{D, \bar D\} = -\partial_x \;,\;\;\; \delta(1,2) = 
\delta(x_1 - x_2)(\theta_1 - \theta_2)(\bar \theta_1 - \bar \theta_2)\;, 
\ee
and the differential operator in the r.h.s. of \p{N2pb} is evaluated 
at the first point of $N=2$ superspace (all derivatives are assumed 
to act freely to the right). 

We observe two differences of $N=2$ case compared to the two previous 
cases. Firstly, $N=2$ supersymmetry requires two fields in the bosonic 
sector, 
the spin $2$ stress-tensor $u(t,x)$ and the spin $1$ current $j(t,x)$ 
which 
generates the affine $\widehat{u(1)}$ subalgebra of $N=2$ SCA. 
The bosonic sector of $N=2$ KdV equation is a coupled system of 
KdV and mKdV equations for these fields. Secondly, there is a 
free parameter $a$ in the hamiltonian and, respectively, in $N=2$ KdV 
equation. 
It was shown [8-10] that this equation is completely 
integrable, i.e. gives rise to an infinite hierarchy of conserved 
hamiltonians in involution and possesses a Lax formulation, only 
for the three special values of $a$ 
\be 
a = -2,\;\; 4, \;\; 1\;. \label{inta}
\ee
These are just the values at which the coupled KdV-mKdV system in 
the bosonic sector of $N=2$ KdV is integrable.

\vspace{0.2cm}
\noindent{\bf 4. N=4 KdV hierarchy.} A natural generalization of $N=2$ 
SCA in the list of Ademollo {\it et al} \cite{Ad} is the 
``small'' $N=4$ SCA. 
Alongside with the conformal stress-tensor, it contains a triplet of the 
spin $1$ currents of affine algebra $\widehat{su(2)}$ and a complex 
$su(2)$ doublet of the spin $3/2$ fermionic currents. It can be formulated 
in a manifestly supersymmetric way as a set of $N=4$ superfield 
PBs \cite{{DI},{DIK}}. 
We will prefer here a $N=2$ superfield notation in which 
this SCA is represented by the $N=2$ stress-tensor $J$ and chiral and 
anti-chiral spin $1$ supercurrents $\Phi$, $\bar \Phi$, $D\Phi = 
\bar D \bar \Phi = 0$ \cite{DIK}. Together with \p{N2pb}, the PBs 
\bea 
\{ J(1), \Phi(2) \} &=& -\left(\Phi \bar D D + 
\bar D\Phi D \right) \delta(1,2)\;,
\{ J(1), \bar \Phi(2) \} = -
\left( \bar \Phi D \bar D + 
D \bar \Phi \bar D \right) \delta(1,2)\; , 
\nonumber \\
\{ \Phi(1), \bar \Phi (2) \} &=& \left(\partial D \bar D 
+DJ \bar D \right)\delta(1,2)\;,\;\;\; \{ \Phi(1), \Phi(2) \} = 0 
\label{N4pb}
\eea
form the classical ``small'' $N=4$ SCA. 

In terms of these supercurrents the transformations promoting  
manifest $N=2$ supersymmetry to $N=4$ are given by
\be
\delta J = -\epsilon \bar D \Phi -\bar \epsilon D \bar \Phi \;, 
\;\; \delta \Phi = \bar \epsilon D J\;, \;\; \delta \bar \Phi = 
\epsilon \bar D J\;.
\label{transform3}
\ee
It is straightforward to check covariance of \p{N2pb}, \p{N4pb} under 
these 
transformations. Then the problem of constructing $N=4$ KdV is reduced 
to constructing most general $N=4$ supersymmetric dimension 
3 hamiltonian 
out of $J, \Phi, \bar \Phi$. It is given by the following expression 
\be
{\cal H}_3 = \int dx d\theta d\bar \theta \left\{J [D,\bar D]J 
- 2\Phi' \bar \Phi + {a\over 6}
J^3 - a J \Phi \bar \Phi - {1\over 2}b \left(\Phi^2 + {\bar \Phi}^2 \right) 
\right\} \label{n4hamn2}
\ee
and contains two real parameters $a$ and $b$, arbitrary for the moment.
The evolution equations for $J, \Phi, \bar \Phi$ can be constructed in 
the standard way, their explicit form can be found in \cite{DIK}. They 
also include the parameters $a$ and $b$.

In ref. \cite{DIK} we investigated the issue of existence of higher-order 
non-trivial conserved hamiltonians for this $N=4$ KdV system and found 
that they exist only for the following three options
\be  \label{opt}
(i).\;\; a = 4, \;b=0;\;\;\;(ii).\;\; a=-2, \;b=6; \;\;\;
(iii).\;\;a=-2, \;b=-6\;.
\ee
Just with these choices the $N=4$ KdV system turns out bi-hamiltonian. 
Both the existence of non-trivial higher-order conserved quantities and 
the bi-hamiltonian property were strong indications that $N=4$ 
KdV system is integrable and gives rise to the whole hierarchy for 
these values of the parameters.

These three choices are essentially different only at first sight. 
Actually, they are related to each other by hidden $SU(2)$ 
symmetry transformations which form an automorphism group of $N=4$, $1D$ 
supersymmetry. The realization of these transformations on  
$N=2$ superfields $J$, $\Phi$, $\bar \Phi$ looks not too illuminating; 
it can be found in ref. \cite{DIK}. 

Both $N=4$ supersymmetry and $SU(2)$ 
covariance become transparent and manifest while formulating the $N=4$ 
KdV system in $N=4$, $1D$ harmonic superspace \cite{DI}. There, the 
``small'' $N=4$ SCA is represented by the analytic doubly-charged 
harmonic superfield $V^{++}(\zeta)$
subjected to the supplementary constraint
\footnote{$\zeta \equiv 
(x, \theta^+, \bar \theta^+, u^+_i, u^-_k)$ are coordinates of an 
analytic subspace of harmonic $N=4$, $1D$ superspace\cite{GIKOS}, 
$u^\pm_i, \;\;
u^{+\;i}u^-_i = 1$ being harmonic coordinates, $\theta^+$, 
$\bar \theta^+$ projections of the $N=4$ grassmann 
coordinates $\theta^i, \bar \theta^k$ on the harmonics $u^+_i$.} 
\be
D^{++}V^{++} = 0, \;\;\;\;\left( D^{++} = u^{+\;i}{\partial 
\over \partial u^{-\;i}} 
+\theta^+\bar \theta^+ \partial_x \right)\;. 
\ee
It restricts the harmonic dependence of $V^{++}$ so that the 
irreducible set of component fields of the latter amounts to 
the currents contents of ``small'' $N=4$ SCA. 
In the ordinary $N=4$ superspace this ``harmonic shortness'' 
condition implies 
$$
V^{++} = V^{(ik)}(x, \theta, \bar \theta) u^+_iu^+_k \;, 
$$
while the analyticity is expressed as the following constraints 
on $V^{(ik)}$ 
$$
D^{(i}V^{kl)} = \bar D^{(i}V^{kl)} = 0\;,
$$
$D^i, \bar{D}^k$ being the appropriate spinor derivatives. The 
automorphism 
$SU(2)$ symmetry in this manifestly $N=4$ supersymmetric 
formulation is realized as rotations of the doublet indices $i,k,l$.

The $N=2$ superfields $J, \Phi, \bar \Phi$ are first components (up to 
numerical coefficients) in 
the decomposition of such $V^{12}, V^{11}$ and $V^{22}$ with respect to 
the grassmann coordinates which enlarge $N=2$ superspace to the $N=4$ one. 

The PB structure \p{N2pb}, \p{N4pb} can be rewritten as a single PB for 
the superfields $V^{++}$ (it is explicitly given in \cite{DI}). 
The hamiltonian \p{n4hamn2}, being expressed through $V^{++}$, takes the 
following form \cite{DI}
\be 
{\cal H}_3 = \int dZ[du] (D^{--}V^{++})^2 + \int d\zeta^{-4}[du]
(a^{--})^2 (V^{++})^3\;. \label{n4hamn4}
\ee   
Here, $D^{--}$ is the second harmonic derivative (not preserving the 
harmonic analyticity), $dZ[du]$ and $d\zeta^{-2}[du]$ are measures of 
integration over the whole harmonic superspace and its analytic subspace, 
the $SU(2)$ breaking parameter $a^{--} = a^{(ik)}u^-_iu^-_k$ is needed 
for balance of the harmonic $U(1)$ charges in the second 
piece of ${\cal H}_3$. Now it is a matter of straightforward though  
tedious calculation to check that the three options \p{opt} just 
correspond to the three (up to reflections) independent orientations 
of the $SU(2)$ breaking constant vector $a^{ik}$ ($(a^{ik})^\dagger 
= -\epsilon_{il}\epsilon_{kt}a^{lt}$)
\bea  
(i).\;\;a^{12} &=& \pm \sqrt{5}\;, \;\;a^{11} = a^{22} = 0\;; \;\;\; 
(ii).\;\;a^{12} = 0\;,  \;\; a^{11} = a^{22} = \pm i \sqrt{5} \;; 
\nonumber\\ 
(iii). \; \; a^{12} &=& 0\;, \;\; a^{11} = - a^{22} = \pm \sqrt{5}\;,
\label{opt1}
\eea
this vector having in all cases the same fixed norm 
\be  \label{intcond}
|a|^2 = -a^{ik}a_{ik} = 2(a^{12}a^{12} - a^{11}a^{22}) = 10\;.
\ee
The latter is the main condition for the $N=4$ KdV system to 
possess an infinite number of higher-order 
hamiltonians in involution and to be bi-hamiltonian \cite{DI}. If 
from the beginning we would keep the central charge $k$ of $N=4$ SCA 
unfixed, in the r.h.s. of eq. \p{intcond} there appeared 
the factor ${1\over k}$. 

In refs. \cite{{IK1},{DG}} two different $N=2$ superfield Lax operators 
for this $N=4$ KdV hierarchy have been proposed. Both of them are 
pseudo-differential and are adapted to the first choice in 
eqs. \p{opt}, \p{opt1}, taking account of the fact that all the three 
options are indeed equivalent by hidden $SU(2)$ covariance. These Lax 
operators are given by       
\bea  
L_1 &=& \partial - J - \bar D \partial^{-1}(DJ) - F\bar D \partial^{-1} 
(D\bar F) + \bar D \partial^{-1}(D(F\bar F)), \;\;
DF = \bar D \bar F = 0\;, \label{lax1} \\ 
(\Phi &=& D\bar F, \;\;\bar \Phi = \bar D F) 
\nonumber \\ 
L_2 &=& D\bar D + D\bar D \partial^{-1}(J + \bar \Phi \partial^{-1} \Phi) 
\partial^{-1}D\bar D\;. 
\label{lax2}
\eea
In both cases the flows and the corresponding conserved hamiltonians are 
given by 
\be
\frac{\partial L}{\partial t_k} = \left[L^k_{\geq 1}, L \right]\;, \;\; 
{\cal H}_n = \int dx d\theta d\bar \theta\; \mbox{res} L^n\;, 
\label{laxfor}
\ee 
the suffix $\geq 1$ meaning the pure differential part 
of pseudo-differential operator.
Note different definitions of the residue of the pseudo-differential 
operators: in the first case it is defined as a coefficient 
before $1$, while in the second case as that before 
$D\bar D\partial^{-1}$. 

A natural reduction to $N=2$ KdV systems is to put in 
\p{lax1} - \p{laxfor}
\be \label{redcond}
\Phi = \bar \Phi =0, 
\ee
which leads to the $a=4$, $N=2$ kdV hierarchy as a consistent reduction 
of the $N=4$ KdV one. All the conserved quantities, as well as the 
above Lax formulations, are reduced to those of this $N=2$ KdV 
hierarchy. However, 
there exists another consistent reduction of $N=4$ KdV. Namely, one 
can choose the $SU(2)$ frame corresponding to the second or third 
options in \p{opt} and also impose the conditions \p{redcond}. 
Though before reductions these options are related to each other by 
the hidden $SU(2)$ symmetry, 
the reductions break $SU(2)$ down to $U(1)$ and so give rise to 
non-equivalent $N=2$ KdV systems. One can show that 
under the second reduction all even-dimensional conserved $N=4$ 
KdV hamiltonians vanish 
(their densities are proportional to $\Phi$, or $\bar \Phi$) while 
the odd-dimensional ones go into those of the $a=-2$, $N=2$ KdV hierarchy. 
For the flows corresponding to these hamiltonians this reduction is 
self-consistent in the sense that both the l.h.s and r.h.s. of 
the evolution equations for $\Phi, \bar \Phi$ vanish upon imposing 
\p{redcond}. Thus two different $N=2$ KdV hierarchy, 
the $a=4$ and $a=-2$ 
ones, are encoded in the single $N=4$ KdV hierarchy as its 
two non-equivalent reductions. The same property can be established 
in the $N=1$ superfield formulation of $N=4$ KdV system \cite{DGI}. 
It would be interesting to find another Lax formulation of $N=4$ KdV, 
such that the existence of these two reductions and the
equivalence of different options in \p{opt} were manifest. 
Hopefully, such a formulation can be constructed in harmonic 
superspace.
 
\vspace{0.2cm}
\noindent{\bf 5. ``Quasi'' N=4 KdV hierarchy.} In ref. \cite{DGI} 
we have found one more integrable hierarchy with the small 
$N=4$ SCA as the second hamiltonian structure. It was naturally 
assigned the name ``quasi'' $N=4$ KdV hierarchy as the global 
$N=4$ supersymmetry 
is explicitly broken down to $N=2$ in this system. Also, 
it reveals no $SU(2)$ covariance and goes over to 
the $a=-2$, $N=2$ KdV upon imposing the conditions \p{redcond}. So it 
can be treated as an integrable extension of this $N=2$ KdV 
hierarchy by chiral and anti-chiral superfields $\Phi$, $\bar \Phi$. 

It is interesting that, at cost of introducing new parameter $c$ (not 
confuse it with the central charge!), the dimension 3 hamiltonian 
of this system (and actually all higher-order hamiltonians) can be written 
uniformly with the $a=4, b=0$ hamiltonian of genuine $N=4$ KdV system  
\be
{\cal H}^c_3 = \int dx d\theta d\bar \theta 
\left\{ J [D,\bar D] J -{c-3\over 3} J^3 - 4J \Phi \bar \Phi 
-2c \Phi'\bar \Phi \right\}\;.
\ee
At $c=1$ the hamiltonian \p{n4hamn2} with $a=4, b=0$ is reproduced 
while at $c=4$ one gets 
the quasi $N=4$ KdV system which goes into the $a=-2$, 
$N=2$ KdV hierarchy 
upon the reduction \p{redcond}. Lacking $N=4$ supersymmetry can be 
easily observed 
already at the level of linear pieces of the corresponding evolution 
equations 
\be
\dot J = - J''' + ...,\;\; \dot \Phi = -c \Phi'''+ ..., \;\; 
\dot{\bar \Phi} = -c \bar{\Phi}''' + ...\;.
\ee
Since $N=4$ supersymmetry \p{transform3} linearly transforms 
$J$, $\Phi$ and $\bar \Phi$ through each other, 
$N=4$ supercovariance strictly requires $c=1$ in these equations. 
So, the case $c=4$ clearly corresponds to the situation 
with broken $N=4$ supersymmetry. 

In \cite{{DGI},{DG}} a scalar Lax formulation for this hierarchy has been 
constructed 
\be
L = D\left( \partial + J - \Phi \partial^{-1} \bar \Phi 
\right) \bar D\;, \;\;\; \frac{\partial L}{\partial t_k} = 
\left[ L^{k/2}_{\geq 1}, L\right]\;. 
\ee
Note that, like the $a=-2$, $N=2$ KdV, this system admits also 
a matrix Lax formulation along the lines of ref. \cite{InK}.

An interesting property of this new $N=2$ hierarchy is that it gives 
rise, via consistent reductions, to two new lower-supersymmetry 
hierarchies with $N=2$ SCA as the second hamiltonian 
structure. They were missed in the previous studies. One of 
them possesses only $N=1$ global supersymmetry and no any kind of 
internal symmetry. The other 
possesses $U(1)$ symmetry but lacks supersymmetry. It is still 
different from the non-supersymmetric system constructed in 
\cite{Mat4}: it 
contains the mKdV hierarchy for the spin $1$ current $j(t,x)$ in its 
bosonic sector, while in the system of ref. \cite{Mat4} this field 
satisfies the trivial equation, $\dot j = 0$. These observations 
suggest the existence of a 
``horizontal'' sequence of hierarchies associated with the given 
SCA. It is parametrized by an extra parameter $c$ which takes, 
similarly to the parameters $a, b$, some special values for 
the integrable cases. These systems range from the maximally 
supersymmetric one to lower-supersymmetric and even 
non-supersymmetric hierarchies. This conjecture implies that 
$N=4$ SCA can serve as the second hamiltonian structure for 
more hierarchies, e.g., respecting only $N=1$ supersymmetry or 
having no supersymmetry at all.  

\vspace{0.2cm}
\noindent{\bf 6. N=4 NLS-mKdV hierarchies.} There exists a remarkable 
and well-known relation between (super)affine and (super)conformal 
algebras: the latter can be mapped on the former through various
Sugawara-Feigin-Fuks (SFF) or coset constructions of (super)conformal 
stress-tensors in terms of the (super)affine currents. Being translated 
into the language of integrable hierarchies, this correspondence 
manifests itself as the relation between two types of hierarchies: 
the KdV type ones associated 
with (super)conformal algebras as the second hamiltonian structure and 
the mKdV type ones which are hierarchies of evolution equations for 
the (super)affine currents with the (super)affine  algebra 
as the hamiltonian structure. In this setting, the SFF 
representations for the stress-tensors come out as Miura-type maps 
between these two sorts of hierarchies. 

Let us again apply to the KdV example. Introduce a
spin $1$ current $v(x)$ generating $\widehat{u(1)}$ affine algebra 
through the PB
\be
\{ v(x), v(y) \} = \partial_x \delta(x-y)\;. \label{u1aff}
\ee
Then, defining 
\be 
u = v^2 +v{}', \label{miura}
\ee
one observes that, as a consequence 
of PB \p{u1aff}, the so defined $u$ generates a classical Virasoro algebra 
\be
\{ u(x), u(y) \} = [ -\partial^3 + 4u\partial + 2u{}'] \delta(x-y)\;, 
\label{conf2}
\ee
which is the same as in eq. (2). Thus eq. \p{miura} gives the simplest 
example of SFF construction relating Virasoro algebra to 
the affine algebra $\widehat{u(1)}$. On the other hand, 
substituting \p{miura} into the 
KdV hamiltonian in (2), one gets 
\be 
{\cal H}_3 = {1\over 2} \int dx \left( v^4 + v{}'v{}'\right)\;.
\ee
Through the PB \p{u1aff} this hamiltonian gives rise to the evolution 
equation for $v$ 
\be 
\dot v = \{ v, {\cal H}_3 \} = -v{}'''- 6 v{}'v^2 \label{mkdv}
\ee
which is the familiar mKdV equation. One can directly check that 
eq. \p{mkdv} yields the standard KdV equation for $u$ defined  
by eq. \p{miura}. Thus the SFF representation \p{miura} is at 
the same time the Miura map relating KdV and mKdV hierarchies. 

This correspondence more or less directly extends to the case of 
supersymmetric hierarchies. E.g., it is easy to check that an $N=2$ 
superextension of the algebra $\widehat{u(1)}$ 
\be
\{ H(1), \bar H(2) \} = D\bar D \delta(1,2), \; \{ H(1), H(2) \} = 0, \; 
DH = \bar D \bar H = 0\;, 
\label{n2haff}
\ee
$H, \bar H$ being spin $1/2$ fermionic chiral and anti-chiral 
superfields (actually, it collects two algebras $\widehat{u(1)}$ 
in its bosonic sector), 
yields just the $N=2$ SCA (11) via the SFF construction 
\be  
J = H\bar H + D\bar H + \bar D H\;. \label{n2miura}
\ee
After substituting this expression into the hamiltonian \p{n2ham}, 
one gets 
a set of evolution equations for $H$, $\bar H$ which give rise to $N=2$ 
mKdV hierarchies for the values of the parameter $a$ listed in eq. 
\p{inta}. For $J$ defined by eq. \p{n2miura} one gets just the related 
$N=2$ KdV hierarchies. 

One can ask whether analogous underlying affine hierarchies can be 
found for $N=4$ KdV and ``quasi'' KdV hierarchies. The answer is 
affirmative, though the proof is not straightforward. 

First of all, it is clear that the relevant superaffine algebras 
should reveal some $N=4$ structure. At present, explicit superfield 
constructions of superextensions of affine algebras and superalgebras 
exist up to $N\geq 1$ ($N$ is as before the number of independent $1D$ 
supercharges). 
In particular, $N=2$ extensions exist for any affine (super)algebra 
admitting a complex structure \cite{HS}. It is natural to assume that  
a hidden $N=4$ supersymmetry is inherent in those $N=2$ affine 
(super)algebras which possess a quaternionic structure, namely those 
which contain as their local part the algebras listed 
in \cite{belg} (actually, this list can be readily extended to 
superalgebras). Then an $N=2$ extension of two affine algebras 
$\widehat{u(1)}$ could be the simplest algebra of this sort. 
It contains in the bosonic sector just four copies of the 
$\widehat{u(1)}$ algebras, the set on which one can already define 
a quaternionic structure \cite{belg}. This $N=2$ algebra is generated 
by two pairs of chiral and anti-chiral superfields 
$H_\alpha, \bar{H}_\alpha$, $(\alpha=1,2)$ with the following PBs 
\be
\{ H_\alpha(1), H_\beta(2) \} = 0\;, \;\;\;\;  \{ H_\alpha(1),  
\bar H_\beta(2) \} = 
\delta_{\alpha\beta} D \bar D \delta(1,2)\;. \label{u111} 
\ee
Indeed, it is easy to see the covariance of these relations, as well as 
of the chirality conditions for $H_\alpha, \bar H_\alpha$, under the 
transformations 
\begin{eqnarray}
\delta H_1 = \epsilon D \bar H_2, \;\; 
\delta \bar H_1 = \bar \epsilon \bar D H_2, \;\; 
\delta H_2 = - \epsilon D \bar H_1,\;\; 
\delta \bar H_2 = - \bar \epsilon \bar D H_1\;. 
\label{transform2}
\end{eqnarray}
They possess the same Lie bracket structure as \p{transform3} and 
so, together with the manifest $N=2$ supersymmetry transformations, 
yield a representation of the same $N=4$ supersymmetry. Hence, the 
above $N=2$ affine algebra indeed supplies the simplest example of 
$N=4$ affine algebra (its supercurrents form 
an $N=4$ supermultiplet).
 
One may wonder whether it gives rise to ``small'' $N=4$ SCA via 
some SFF construction and has any relation to $N=4$ KdV hierarchy. 
However, it can be checked that one cannot construct, out of the 
superfields $H_\alpha, \bar H_\beta$, any $N=4$ multiplet of composite 
currents including $N=2$ conformal stress-tensor with 
a Feigin-Fuks term (the latter is absolutely necessary for 
producing a central term in $N=2$ SCA and thus generating at least 
an $N=2$ KdV hierarchy as a subsystem). The only possibility is the 
$N=4$ multiplet 
\be
\hat{J} = H_1\bar H_1 + H_2 \bar H_2,\;\;\; \hat{\Phi} 
= H_1H_2, \;\; \hat{\bar \Phi} = \bar H_2 \bar H_1\;, 
\ee
which, via PBs \p{u111}, generates a topological (i.e. centreless) 
``small'' $N=4$ SCA. So, possible $N=2$ affine hierarchies 
(even possessing rigid $N=4$ supersymmetry) constructed on the basis 
of this $N=2$ affine algebra seem to have 
no any direct relation to $N=4$ KdV hierarchy. 

Next in complexity is $N=2$ extension of non-abelian affine algebra 
$\widehat{sl(2)\oplus u(1)}$ whose local bosonic part 
$sl(2)\oplus u(1)$ also contains four generators and is among 
the algebras given in \cite{belg}. 
The PBs of these $N=2$ superalgebra read \cite{HS} 
\begin{eqnarray}
\{ H(1), \bar H(2) \} &=& D \bar D \delta (1,2)
\label{hhbr} \\
\{ H (1) , F(2) \} &=& D F \delta (1,2)\; , 
\;\;\;
\{ H (1) , \bar  F (2) \} =
-D \bar F \delta (1,2)\; , \nonumber\\
\{ \bar H (1) , F(2) \} 
&=&- \bar D F \delta (1,2) \; , \;\;\;
\{ \bar H (1) , \bar F (2) \} =
 \bar D  \bar F  \delta (1,2)\;, 
\label{u2al2}\\
\{F(1), \bar F (2) \}&=& 
\left[ (D + H)( \bar D + \bar H ) + 
F \bar F \right] \delta (1,2)\; ,
\label{u2al3}
\end{eqnarray}
all other PBs vanishing. Here, $H$ and $\bar H$ satisfy the standard 
chirality conditions while $F$ and $\bar F$ are subject to the 
{\it nonlinear} version of chirality  
\be 
\left( D  + H\right)F = 0,\;\;\left( \bar D -
\bar H \right) \bar F =0 \;. \label{covchir}
\ee
These constraints are necessary for closure of Jacobi identities of the 
algebra \cite{HS}. 

Once again, it is a matter of direct calculation to check that these 
PBs together with the above linear and nonlinear chirality conditions 
are covariant under the following hidden {\it nonlinear} 
$N=2$ transformations 
\cite{IKT} 
\begin{eqnarray}
\delta H &=& \epsilon D \bar F + \bar \epsilon H F,\;\;\;
\delta \bar H = \bar \epsilon \bar D F - \epsilon
\bar H \bar F \nonumber\\
\delta F &=& - \epsilon D \bar H - \epsilon ( H \bar H
+ F \bar F ), \;\;\;
\delta \bar F = - \bar \epsilon \bar D H -
\bar \epsilon  ( H \bar H + F \bar F).
\label{transform}
\end{eqnarray}
It can be easily checked that the above transformations, despite their 
non-linearity, indeed realize the same extra $N=2$ supersymmetry as 
the transformations \p{transform3}. Thus, combined with manifest $N=2$ 
supersymmetry, they again form the previously defined $N=4$ 
supersymmetry. 
The two pairs of affine $N=2$ supercurrents 
$F, \bar F$ and $H, \bar H $ are unified into 
an irreducible $N=4$ supermultiplet, so the $N=2$ extension of 
$\widehat{sl(2)\oplus u(1)}$ algebra is in fact an $N=4$ extension. 
This $N=4$ structure might be made manifest by passing to $N=4$ 
superfields, but we will not elaborate on this possibility here. 

What is indeed important for our consideration is that this superaffine 
algebra allows for a transparent SFF construction of small $N=4$ SCA on 
its basis. The explicit formulas expressing $N=4$ supercurrents in terms 
of the affine supercurrents are as follows \cite{IKT}
\be
J = H \bar H + F \bar F + D \bar H 
+ \bar D H \;,\;\;\;\Phi = D \bar F\;,\;\;\; \bar \Phi =
\bar D F\;, \;\;( \;D \Phi =  \bar D \bar \Phi = 0\;). 
\label{N2stress}
\ee
These objects obey just the $N=4$ SCA PB relations \p{N2pb}, 
\p{N4pb} as a 
consequence of the affine PB relations \p{hhbr} - \p{u2al3}. 
Due to the Feigin-Fuks term in $J$ in \p{N2stress} the resulting 
$N=4$ SCA possesses a 
non-zero central charge, which, as was already mentioned, is crucial 
for getting $N=4$ KdV system. 

Now it is a standard routine to substitute these SFF expressions 
into the hamiltonians of $N=4$ KdV hierarchy and to derive the flow 
equations for the affine currents $H, \bar H, F, \bar F$ using the 
PBs of $N=2$ affine $\widehat{sl(2)\oplus u(1)}$ algebra. Note that the 
lowest flow equations were derived in \cite{IKT} also in another way,  
by the direct construction of the proper dimension $N=4$ invariant 
hamiltonians and requiring them to be in involution. We do not present 
here these equations in view of their considerable complexity (see ref. 
\cite{IKT}). We only notice the existence of a few consistent reductions 
of them.

The first one is effected by putting 
\be \label{red1}
F =\bar F =0 \rightarrow \Phi = \bar \Phi =0, \; J = H\bar H +D\bar H + 
\bar D H\;. 
\ee
This yields the $a=4$, $N=2$ mKdV \footnote{Using another choice of 
the frame with respect to the hidden automorphism $SU(2)$, it is 
possible to perform a reduction to the $a=-2$, $N=2$ mKdV 
hierarchy as well.}.

The second reduction goes as 
\be \label{red2}
H = \bar H = 0 \rightarrow J = F\bar F, \;\; DF = \bar D \bar F =0\;.
\ee
The resulting system is the $N=2$ NLS hierarchy of refs. \cite{nls}. 
The existence of such a reduction has been firstly noticed in \cite{IK1} 
at the level of $N=4$ KdV hierarchy, with $F$ and $\bar F$ 
interpreted as the prepotentials of the spin $1$ chiral supercurrents 
$\Phi$ and $\bar \Phi$ in a fixed gauge with respect to the prepotential 
gauge freedom. 

This consideration shows that the $N=4$ supersymmetric system constructed 
can be treated as $N=4$ extension of at once two $N=2$ supersymmetric 
hierarchies, $N=2$ mKdV and NLS ones. This is why it has been named in 
\cite{IKT} the ``$N=4$ NLS-mKdV hierarchy''.

Needless to say, the ``quasi'' $N=4$ KdV hierarchy can also 
be associated with some underlying $N=2$ $\widehat{sl(2)\oplus u(1)}$ 
affine hierarchy. One simply substitutes the STT expressions \p{N2stress} 
for the $N=4$ supercurrents into the hamiltonians of the ``quasi'' $N=4$ 
KdV hierarchy and Lax operator and derives the appropriate evolution 
equations for the affine supercurrents via the PB structure 
\p{hhbr} - \p{u2al3}.

\vspace{0.2cm}
\noindent{\bf 7. Concluding remarks.} Thus, now we are aware of 
general method of setting up $N=4$ supersymmetric KdV type 
hierarchies: each $N=2$ affine algebra or superalgebra admitting a 
hidden $N=4$ supersymmetry (quaternionic structure) can be used to 
construct such hierarchies via appropriate superfield SFF maps. We hope 
to naturally come in this way to $N=4$ extensions of $W$ algebras. 
An example of $N=2$ affine 
algebra with $N=4$ structure, next in complexity to $N=2$ 
$\widehat{sl(2)\oplus u(1)}$, is the algebra $N=2$ $\widehat{sl(3)}$. 
This case is under study. 

There remain many conceptual and technical problems to be solved. 
In particular, it would be useful to work out convenient $N=4$ superfield 
techniques  of treating $N=4$ hierarchies, based, e.g., on the harmonic 
superspace approach. Up to now, it has been successfully applied only to 
one example of $N=4$ KdV type hierarchies, the $N=4$ KdV system 
itself \cite{{DI},{DIK}}. An interesting unsolved problem is to construct 
super KdV hierarchy associated with the ``large'' $N=4$ SCA as the second 
hamiltonian structure. There are indications that such a system could yield 
all the three known $N=2$ supersymmetric KdV hierarchies as its 
different consistent 
reductions. One may also think about higher $N$ hierarchies, say, with 
$N=8$ supersymmetry. To my knowledge, no any ``no-go'' theorems are known 
which could forbid the existence of such systems. Also, it could happen 
that a number of the already known $N=2$ hierarchies exhibit hidden 
higher $N$ supersymmetries. For instance, in a recent preprint \cite{sor} 
it was found that $N=4$ KdV hierarchy allows a map on the so called 
$(1,1)$ GNLS (Generalized NLS) system \cite{BKS} which involves one chiral 
fermionic and one chiral bosonic superfields. 

Perhaps, the most urgent problem is to identify the place 
and to reveal possible implications of this novel wide class of 
supersymmetric integrable systems in the modern superstring 
and $p$- brane stuff. I believe this certainly can be done. 

\vspace{0.3cm}
\noindent{\bf Acknowledgement.} I am grateful to Organizers of 
the D.V. Volkov Memorial Seminar for inviting me to give this Talk. 
I thank Loriano Bonora for hospitality at S.I.S.S.A., Trieste, where 
this work was finalized. A partial support from RFBR grant 
RFBR 96-02-17634, INTAS grant INTAS-94-2317 and a grant of the 
Dutch NWO organization is acknowledged.

\end{document}